\documentclass[journal]{IEEEtran}

\usepackage{cite}

\usepackage[pdftex]{graphicx}
\DeclareGraphicsExtensions{.pdf,.jpeg,.png}

\usepackage{amsmath}

\usepackage[hidelinks]{hyperref}
\usepackage[dvipsnames]{xcolor}
\usepackage{amssymb}
\usepackage{mathrsfs}
\usepackage[normalem]{ulem}
\usepackage{enumitem}
\usepackage{tabularx}
\usepackage{makecell}
\usepackage{multirow}
\usepackage{ulem}
\usepackage[flushleft]{threeparttable}
\usepackage{gensymb}
\usepackage{subcaption}
\usepackage{cleveref}[2012/02/15]
\crefformat{footnote}{#2\footnotemark[#1]#3}

\def\x{\boldsymbol{x}}

\def\v{\boldsymbol{v}}

\begin{document}

\title{Towards Robust Sensing for Autonomous Vehicles: An adversarial perspective}

\author{
    \IEEEauthorblockN{Apostolos Modas\IEEEauthorrefmark{2}\IEEEauthorrefmark{1}, Ricardo Sanchez-Matilla\IEEEauthorrefmark{4}\IEEEauthorrefmark{1}, Pascal Frossard\IEEEauthorrefmark{2}, Andrea Cavallaro\IEEEauthorrefmark{4}}\\
    \IEEEauthorblockA{\IEEEauthorrefmark{4}Centre for Intelligent Sensing, Queen Mary University of London, UK}\\
    \IEEEauthorblockA{\IEEEauthorrefmark{2}{Ecole Polytechnique F\'ed\'erale de Lausanne (EPFL), Switzerland}}
\thanks{\IEEEauthorrefmark{1}Equal contribution. \newline \copyright~2020 IEEE. Personal use of this material is permitted.  Permission from IEEE must be obtained for all other uses, in any current or future media, including reprinting/republishing this material for advertising or promotional purposes, creating new collective works, for resale or redistribution to servers or lists, or reuse of any copyrighted component of this work in other works.}
}




\maketitle



\IEEEpeerreviewmaketitle

\IEEEPARstart{A}{utonomous} Vehicles rely on accurate and robust sensor observations for safety critical decision-making in a variety of conditions. Fundamental building blocks of such systems are sensors and classifiers that process ultrasound, RADAR, GPS, LiDAR and camera signals~\cite{Khan2018}. It is of primary importance that the resulting decisions are robust to perturbations, which can take the form of different types of nuisances and data transformations, and can even be adversarial perturbations (APs). 

Adversarial perturbations are purposefully crafted alterations of the environment or of the sensory measurements, with the objective of attacking and defeating the autonomous systems. A careful evaluation of the vulnerabilities of their sensing system(s) is necessary in order to build and deploy safer systems in the fast-evolving domain of AVs. To this end, we survey the emerging field of sensing in adversarial settings: after reviewing adversarial attacks on sensing modalities for autonomous systems, we discuss countermeasures and present future research directions.

\section{Robust Sensing}
\label{subsec:definitions}

The robustness of a decision system refers to its capability of making the correct decision, even when testing conditions are degraded. In particular, it is important that the decisions are not changed when small perturbations alter the input signal. The lack of robustness poses critical safety threats for AVs, and it could even allow an attacker to design adversarial perturbations that defeat their sensing systems.


In order to formalize the concept of APs, let $f_m$ be a decision-making system (e.g.,~a classifier or detector) for a sensing modality $m$, $\x$ an input signal of the same modality belonging to a data distribution $\mathcal{X}$, $g_{\v}$ a function that applies a perturbation $\v$ on the input signal $\x$, and $\mathcal{D}$ a function that measures the distortion between the input and the perturbed signal. Then, $\v$ is called an AP, and consequently, $g_{\v}(\x)$ is an adversarial attack that generates adversarial examples, if:
\begin{equation}
\begin{aligned}
f_m\big(g_{\v}(\x)\big) \neq f_m(\x), \\
\text{subject to} \quad \mathcal{D}\big(g_{\v}(\x), \x\big) < \epsilon, \\
g_{\v}(\x) \in \Phi,
\end{aligned}
\label{eq:def_adversarial_perturbation}
\end{equation}
where $\Phi$ is the legitimate domain of the original signal $\x$, and $\epsilon$ limits the distortion of the input signal. This limitation relates to the most common definition of adversarial attacks, and implies that the information of the sensed signal is not considerably changed, or even destroyed, by an attack. 
For example, it is common to control the  distortion with the constraint that the original (clean) and the adversarial signals should be close enough in an $\ell_p$-norm sense, \mbox{$\| g_{\v}(\x) - \x\|_p < \epsilon$}.
While several adversarial methods focus on additive APs, where $g_{\v}(\x)=\x+\v$, the adversarial attack $g_{\v}$ can be any function that computes and applies a perturbation on a signal (e.g.,~transformation-based APs).

\begin{figure}[t!]
    \centering
    \includegraphics[width=0.9\columnwidth]{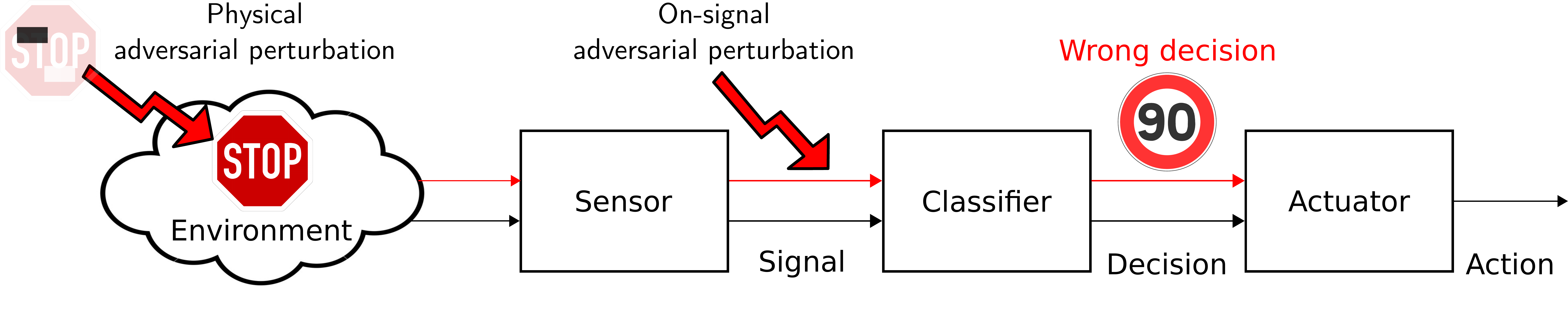}
    \caption{The physical and on-signal adversarial attacks.}
    \label{fig:concept}
\end{figure}

In general, adversarial attacks can be {physical} or {on-signal} (see  Figure~\ref{fig:concept}). {\em Physical adversarial attacks} are perturbations that modify the environment in an adversarial manner, for example, to mislead the interpretation of traffic signs~\cite{Eykholt2018}. Physical attacks can hide objects (i.e.,~causing false-negative detections), generate observations of inexistent objects (i.e.,~producing false-positive detections), induce misclassification of detected objects, or counterfeit signals. Physical APs are generally constructed under transformation-based constraints, to be invariant to transformations that take place when captured in the physical world, such as the rotations, scaling, and illumination variations that can occur in cameras, for example. {\em On-signal adversarial attacks} modify the data captured by sensors within the decision system. These attacks, which expose the vulnerabilities of the classifiers or detectors, have been mostly used to evaluate, understand, and improve decision systems~\cite{Fawzi2016}, \cite{Fawzi2017}, \cite{ ortiz-jimenezHoldMeTight2020}. However, the ability to design an on-signal attack implies that the attacker already has {complete access} to the system and to the captured signal. As this scenario relates to the (cyber-) security of the AV, it is out of the scope of this article. An extensive overview of on-signal attacks is provided in~\cite{Yuan2019}.

We focus now on physical attacks that alter the behavior of AVs, exclusively by external actions. For example, attacks that {\em hide objects} (jamming) may prevent AVs from detecting pedestrians, or possibly cause a collision. Attacks that add {\em objects} (spoofing) force AVs to detect inexistent objects, possibly causing the AVs to suddenly stop. {\em Misclassification} attacks aim to change the decision of a classifier, such as interpreting a stop sign as speed limit sign, thus inducing wrong actions. Finally, {\em counterfeiting} attacks mimic the properties of the original signals with the aim of delivering adversarial measurements to a sensor (e.g.,~delayed or wrong GPS signals). 

\begin{figure*}[t!]
    \centering
    \setlength\tabcolsep{3pt}
    \begin{tabular}{ccc}
        \includegraphics[height=4.4cm,trim={0cm 0 0cm 0},clip]{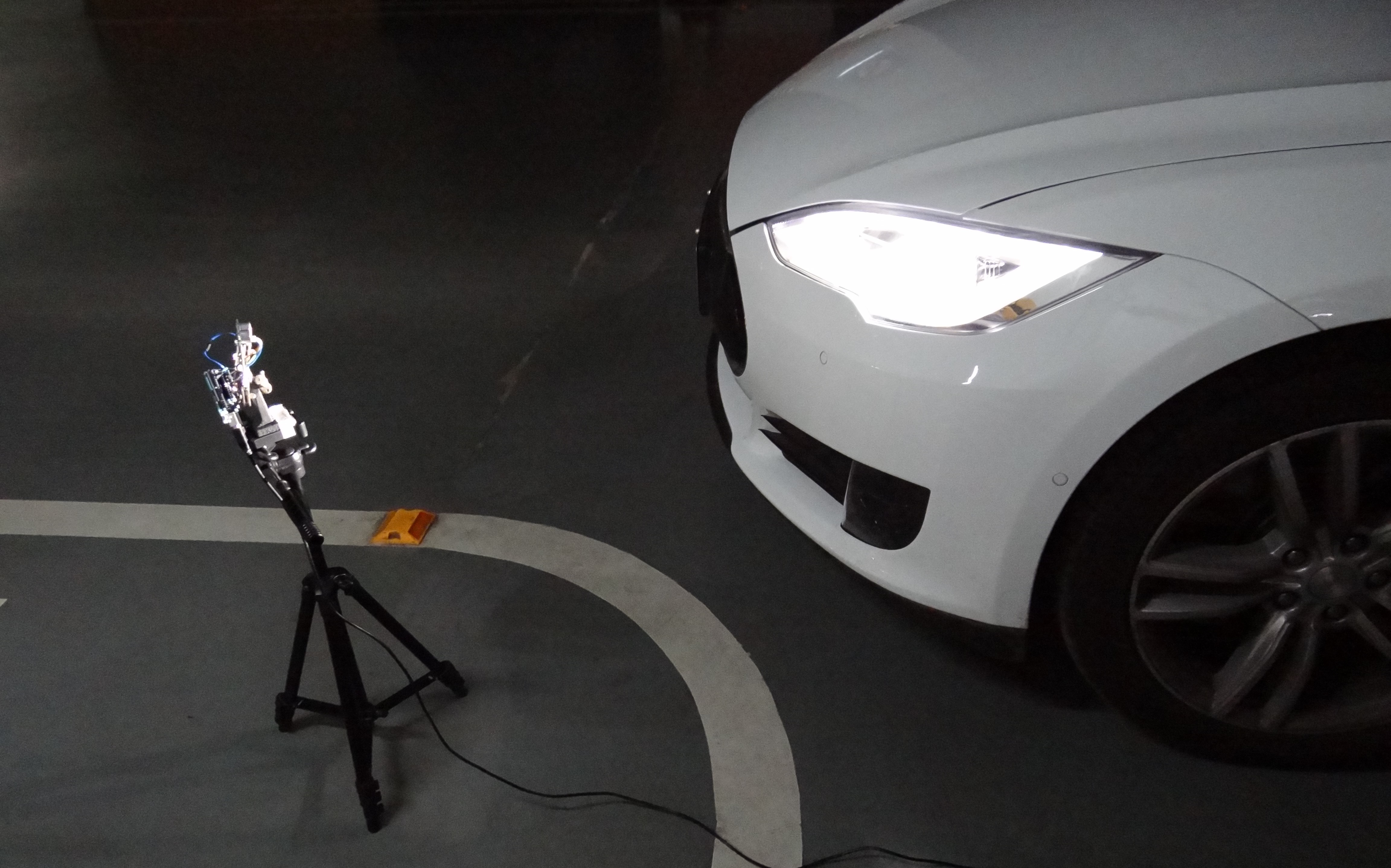} &
        \includegraphics[height=4.4cm]{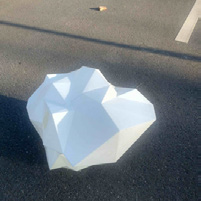} &
        \includegraphics[height=4.4cm]{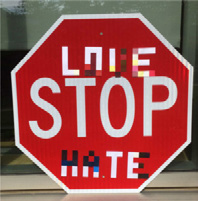} \\
        (a) Device & (b) Object & (c) Patch 
    \end{tabular}
\caption{Examples of the three \emph{types} of adversarial attacks: (a) with a spoofing \emph{device} to hide objects from an ultrasonic sensor~\cite{yan2016},
(b) with an adversarial \emph{object} that is not detectable by LiDAR~\cite{Cao2019a}, and
(c) with a \emph{patch} on a traffic sign that causes misclassification with cameras~\cite{Eykholt2018}. Images provided by the authors.}
\label{fig:attack_types}
\end{figure*}

\begin{figure*}[t!]
    \centering
    \setlength\tabcolsep{3pt}
    \begin{tabular}{ccc}
        \includegraphics[height=4.5cm,trim={6.5cm 0.5cm 7cm 0.5cm},clip]{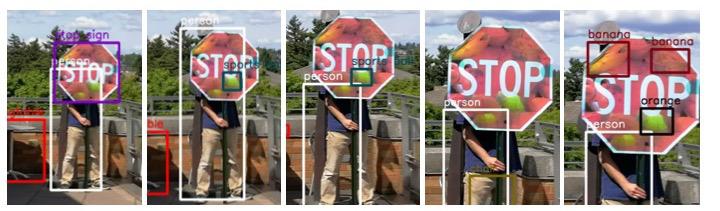} &
        \includegraphics[height=4.5cm,trim={0.5cm 0cm 0.5cm 0cm},clip]{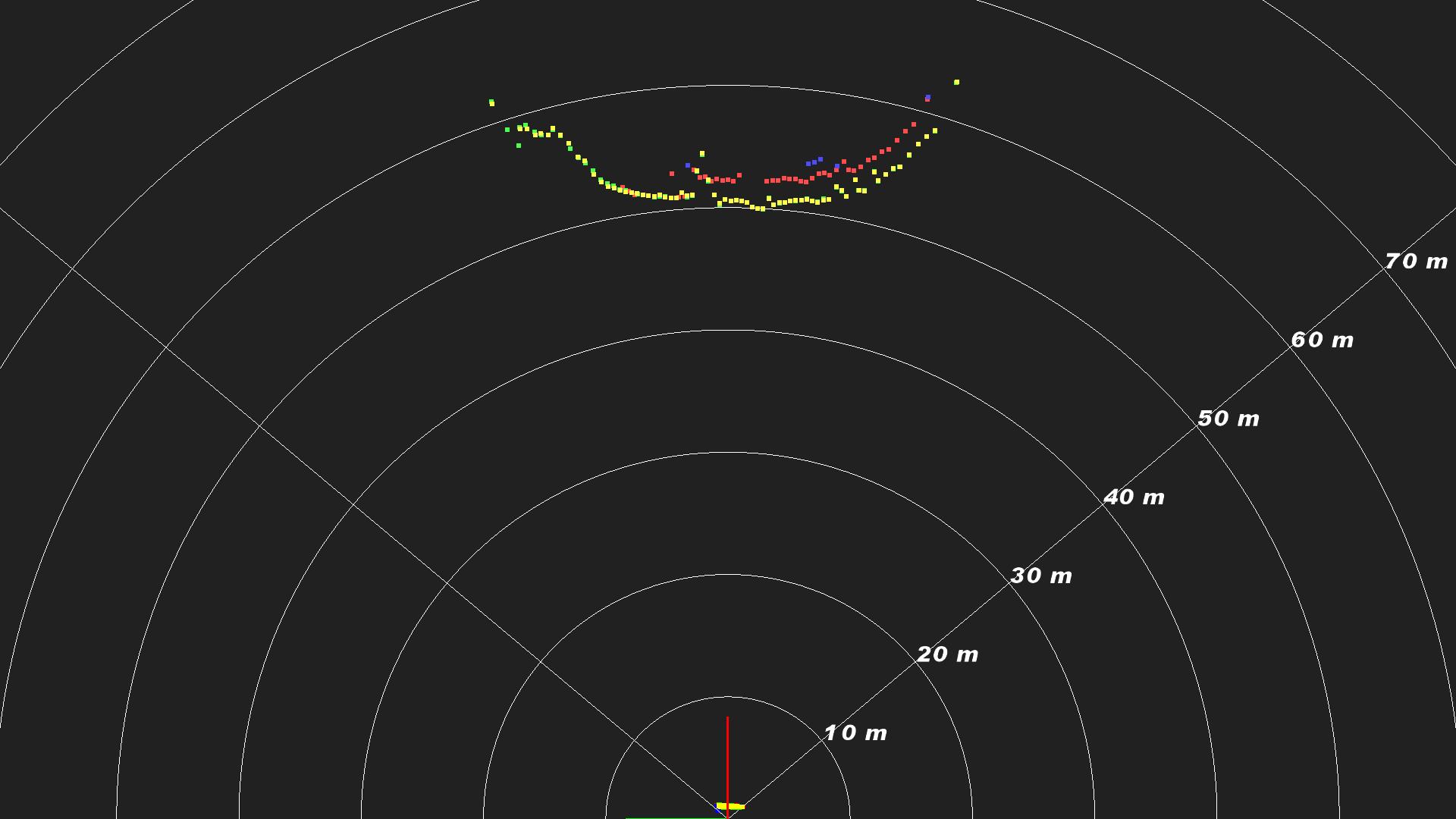} &
        \includegraphics[height=4.5cm]{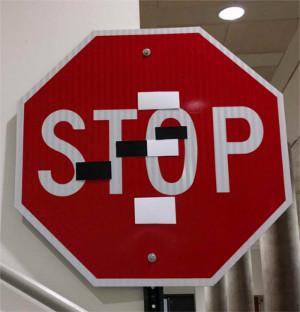} \\
        (a) Hiding & (b) Adding & (c) Misclassifying
    \end{tabular}
\caption{Examples of three \emph{objectives} for adversarial attacks: (a) sign that is not visible by a camera (\emph{hidden object})~\cite{Eykholt2018b},
(b) with a spoofing attack that \emph{adds} nonexistent obstacles for LiDAR~\cite{Petit2015RemoteAO}, and
(c) with patches on a traffic sign that causes a camera to \emph{misclassify} the traffic sign~\cite{Eykholt2018}. Images provided by the authors.}
\label{fig:attack_goal}
\end{figure*}

We can distinguish attacks based on the means used by the attackers. Physical adversarial attacks can be perpetrated using devices, objects, or patches (see Figure~\ref{fig:attack_types}). \emph{Devices} (e.g.,~a laser pointer) directly target the sensor. An adversarial \emph{object}, such as an adversarial traffic sign, can be constructed to mislead a classifier, even if its appearance may be similar to that of a benign object. Finally, a \emph{patch}, such as a crafted sticker, can be placed on objects in the environment to mislead classifiers.

Next, depending on the attacker's knowledge of the underlying system, attacks can be categorized as white, gray, or black-box. In \emph{white-box} settings, the attacker has complete knowledge of the system (e.g.,~its architecture and parameters). When machine learning models are used, the attacker also knows the training data used to generate the model. In \emph{black-box} settings, the attacker has access only to the output of the system. Finally, in \emph{gray-box} settings, the attacker has only partial information of the system, such as the architecture but not the parameters or the training data, or the architecture and the training data but not the parameters.


APs can also be categorized as {data-specific} or {data-agnostic}. A \emph{data-specific} AP is a perturbation $\v$ computed for a \emph{specific} signal $\x$ of a sensing modality~\cite{Sitawarin18}, and it can efficiently apply to that particular signal only (e.g.,~a specific stop sign). An attacker can also construct \emph{data-agnostic} APs, commonly referred to as \emph{universal} APs~\cite{Seyed-Mohsen16}, which are applicable to \emph{any} signal, $\forall\x\sim\mathcal{X}$, of a specific modality. An example here is a single perturbation for any street sign that deceives a sensing system with high probability~\cite{Eykholt2018}. 
A subset of data-agnostic perturbations, called \emph{class-universal}, are data-agnostic only for intra-class signals~\cite{Thys19} (i.e.,~a single perturbation for every stop sign); that is, $\forall\x\sim\mathcal{X}_c$, where $c$ is the corresponding class. {Data-agnostic and class-universal APs are critical for the robustness of AVs}, since a single AP added to the physical world could potentially mislead \emph{any} classifier of a specific modality, thus posing a safety threat. This \emph{potential} threat derives from the fact that a  strong property of data-agnostic and class-universal APs is their \emph{transferability}:  APs computed on a known model can also mislead an unknown model~\cite{Seyed-Mohsen16}.

In the next sections, we analyze physical adversarial attacks and how they affect different sensors. We characterize each attack based on its type, goal (Figure~\ref{fig:attack_goal}), data-dependency, system knowledge, and evaluation. To evaluate the attacks, we use the term \emph{attack success rate} to denote the percentage of adversarial examples that successfully mislead the system. State-of-the-art methods are compared in Table~\ref{tab:soa_attacks} using the proposed taxonomy.

\begin{table*}[t]
\centering
\setlength\tabcolsep{2pt}
\caption{Physical adversarial attacks per sensing modality. KEY: H: hiding objects; A: adding objects; M: misclassifying objects; C: counterfeit signal; DS: data-specific; CU: class-universal; DA: data-agnostic; WB: white-box; GB: gray-box; BB: black-box; AVs: evaluation on autonomous vehicles settings. Note that~\cite{Papadimitratos2008b} is evaluated only on simulations.}
\begin{tabular}{ll|ccc|cccc|ccc|ccc|cc}
\Xhline{3\arrayrulewidth}
\multicolumn{1}{c}{\multirow{2}{*}{\textbf{Sensor}}} & \multirow{2}{*}{\textbf{Reference}} & \multicolumn{3}{c}{\textbf{Type}} & \multicolumn{4}{c}{\textbf{Goal}} & \multicolumn{3}{c}{\textbf{Data dependency}} & \multicolumn{3}{c}{\textbf{Knowledge}} & \multicolumn{2}{c}{\textbf{Evaluation}} \\
\multicolumn{1}{c}{} & & \textbf{Device} & \textbf{Object} & \textbf{Patch} & \textbf{H} & \textbf{A} & \textbf{M} & \textbf{C} & \textbf{DS} & \textbf{CU} & \textbf{DA} & \textbf{WB} & \textbf{GB} & \textbf{BB} & \textbf{Physical} & \textbf{AVs} \\
\Xhline{3\arrayrulewidth}
\multirow{2}{*}{\textbf{Ultrasonic}} & \cite{yan2016} & \checkmark &  &  & \checkmark & \checkmark &  &  &  &  & \checkmark &  &  & \checkmark & \checkmark & \checkmark \\ 
 & \cite{Xu2018} & \checkmark &  &  & \checkmark & \checkmark &  &  &  &  & \checkmark &  &  & \checkmark & \checkmark & \checkmark \\ \hline
\textbf{RADAR} & \cite{yan2016} & \checkmark &  &  & \checkmark &  &  &  &  &  & \checkmark &  & \checkmark &  & \checkmark & \checkmark \\ \hline
\multirow{2}{*}{\textbf{GPS}} & \cite{Humphreys2008} & \checkmark &  &  &  &  &  & \checkmark &  &  & \checkmark & \checkmark &  &  & \checkmark &  \\
 & \cite{Papadimitratos2008b} & \checkmark &  &  &  &  &  & \checkmark & \checkmark &  &  & \checkmark &  &  &  &  \\ \hline
\multirow{4}{*}{\textbf{LiDAR}} & \cite{Petit2015RemoteAO} & \checkmark &  &  & \checkmark & \checkmark &  &  &  &  & \checkmark &  & \checkmark &  & \checkmark &  \\
 & \cite{Shin2017} &  \checkmark &  &  & \checkmark & \checkmark &  &  &  &  & \checkmark & \checkmark &  &  & \checkmark &  \\
 & \cite{Cao2019b} &  \checkmark &  &  &  & \checkmark &  &  &  &  & \checkmark & \checkmark &  &  &  & \checkmark \\
 & \cite{Cao2019a} &   & \checkmark &  & \checkmark &  & \checkmark &  & \checkmark &  & \checkmark & \checkmark &  & \checkmark & \checkmark & \checkmark \\ \hline
\multirow{10}{*}{\textbf{Camera}} & \cite{Petit2015RemoteAO} &  \checkmark &  &  & \checkmark &  &  &  &  &  & \checkmark &  & \checkmark &  & \checkmark &  \\
 & \cite{yan2016} & \checkmark &  &  & \checkmark &  &  &  &  &  & \checkmark &  &  & \checkmark & \checkmark &  \\
 & \cite{Athalye2017} & & \checkmark &  &  &  & \checkmark &  & \checkmark &  &  & \checkmark &  &  & \checkmark &  \\
 & \cite{Sitawarin18} &   & \checkmark &  &  &  & \checkmark &  & \checkmark &  &  & \checkmark &  &  & \checkmark & \checkmark \\
 & \cite{Chen18} &   & \checkmark &  &  &  & \checkmark &  &  & \checkmark &  & \checkmark &  &  & \checkmark & \checkmark \\
 & \cite{Eykholt2018} &   & \checkmark & \checkmark &  &  & \checkmark &  &  & \checkmark &  & \checkmark &  & & \checkmark & \checkmark \\
 & \cite{Eykholt2018b} &  & \checkmark & \checkmark & \checkmark & \checkmark &  &  &  & \checkmark &  & \checkmark &  &  & \checkmark & \checkmark \\
 & \cite{Brown2017} &   &  & \checkmark &  &  & \checkmark &  &  &  & \checkmark & \checkmark &  &  & \checkmark &  \\
 & \cite{Thys19} &   &  & \checkmark & \checkmark &  &  &  &  & \checkmark &  & \checkmark &  &  & \checkmark &  \\
 & \cite{Lee2019} &  &  & \checkmark & \checkmark &  &  &  &  &  & \checkmark & \checkmark &  &  & \checkmark & \\
 \Xhline{3\arrayrulewidth}
\end{tabular}
\label{tab:soa_attacks}
\end{table*}
\section{Attacks using devices}

In this section we discuss physical adversarial methods that use devices to attack ultrasonic sensors, RADAR, GPS, LiDAR and cameras.

\subsection{Ultrasonic and RADAR sensors}

\emph{Black-box} and \emph{data-agnostic} jamming and spoofing attacks for both millimeter-wave radars and ultrasonic sensors aim to \emph{hide} or \emph{add} objects in tasks such as automatic parking or moving in reverse gear~\cite{yan2016}. The jamming transducer generates ultrasonic signals approximately around the sensor's resonant frequency. The jamming signal exceeds the sensor's threshold for the detection of returning echoes (it lowers the signal-to-noise ration), and prevents the sensor from detecting a parked vehicle. The spoofing attack  must be executed in the interval between the end of the transmitted pulse and the start of the first echo; hence, the echoes are injected with a cycle time that is several milliseconds shorter than the sensor's. This  may cause unstable spoofed sensor readings, but it guarantees successful injection in the attack slot.
During parking, these attacks lead to collisions, as objects are successfully hidden. When AVs are moving in reverse gear, these attacks are successful at a distance of up to 10~meters. However, these methods are applicable only to slowly moving vehicles.

\emph{Black-box} and \emph{data-agnostic} attacks for ultrasonic sensors also include random spoofing, adaptive spoofing, and jamming~\cite{Xu2018}. These attacks aim to \emph{hide} or \emph{add} objects. Emitting spoofed signals once every few milliseconds causes the AV to stop moving as it detects inexistent objects. These false positives can only be created at distances closer than the distance between the spoofer device and the vehicle. To create inexistent objects at an arbitrary yet stationary distance, the adaptive spoofer pings a signal, listens and adjusts to the concurrent sensor signals adaptively, eliminates the existing echoes, and transmits the spoofed signal. Similarly to~\cite{yan2016}, the jamming attack continuously emits ultrasound waves around the resonant frequency of the sensors. 
Adaptive spoofing can deceive the AV into detecting a nonexistent object at any distance within the sensor range. Jamming prevents an AV from detecting objects, thus potentially causing collisions during automatic parking.

Finally,   \emph{gray-box} jamming and spoofing attacks can target RADAR sensors to \emph{hide} objects (i.e.,~generate false negatives) or to \emph{add} pseudo-objects (i.e.,~produce false positives)~\cite{yan2016}. Using a signal analyzer to identify the center frequency of the (known) radar sensor, an attacker can generate a jamming waveform, the frequency of which is close to  the center frequency of the sensor, thus preventing the detection of neighbouring vehicles. For the spoofing attack, the adversarial signals are modulated as those of the automotive radar. However, since the spoofing attack must be executed in the interval between the end of the transmitted pulse and the start of the first echo, the slope of the ramp is tuned back and forth in a higher value range on the signal. This produces periodic distance changes in the estimations that lead to false identification of pseudo-objects.
Similar to the attacks on ultrasonic sensors, these attacks are applicable only against  slowly moving vehicles.


\subsection{GPS}
Attacks on GPS sensors usually aim to \emph{counterfeit} the GNSS transmitted signals with an adversarial transmitter that emits signals identical to those sent by the satellites, so that the location computed by the GNSS receivers (i.e.,~GPS) is incorrect. Such attacks are \emph{data-agnostic} and, usually, \emph{white-box}, since the attacker must know the GNSS implementation  of the receivers and transmitters. 

The most common adversarial techniques are spoofing and replay attacks.
An attacker must first transmit a sufficiently powerful jamming GNSS signal to force the receivers to lose their lock on the satellite. Then, the attacker can forge and transmit spoof GNSS signals at the receivers' antennas, at the same frequency and with a higher power than those of the legitimate GNSS signals. If the lock of the attacked receivers on the spoofed signal persists, then the location estimation is under the influence of the attacker. An attacker can spoof a varying number of GPS receivers to any location, while preserving their mutual time offsets. An analysis of the requirements, architecture, and application of a receiver-spoofing attack is given in~\cite{Humphreys2008}.

Replay attacks can be characterized by the capability of the adversarial node to receive, record, and replay GNSS signals, and to the delay time between reception and re-transmission~\cite{Papadimitratos2008b}. Since the attacker can replay recorded signals with any additional delay, the receivers will start receiving the replayed navigation messages after some adversary-specific signal propagation delay, which is usually negligible. The delay,  the most important feature, is used to control the shift in the position, velocity, time (PVT) computation by the receivers. The randomized replay attack can create (in simulations)  location offsets on the order of hundreds of kilometers.

\subsection{LiDAR}
We can identify three types of attacks~\cite{Petit2015RemoteAO}, namely front/rear/side (the attacker is an adversarial vehicle that moves close to the AV), roadside (the attacker installs malicious devices on the roadside), and  mechanic (the attacker has limited time to place a malicious device on the AV). Relay and  spoofing \emph{data-agnostic} and \emph{gray-box} attacks can be used to \emph{hide} or to \emph{add} objects.   Note that these attacks assume that the technical specifications and datasheets of the sensors are publicly known. During the relay attack, the original signal sent from the target vehicle LiDAR is relayed from another position to create fake echoes and make real objects appear closer or farther than their actual locations. The spoofing attack  creates inexistent objects: the original signal is used as a trigger to actively spoof the LiDAR sensor, with the intention to relay or replay objects and control their positions.
The relay attack can hide objects that are 1~meter away, while generating false-positive detections 20--50~meters away. The success of the replay attack is limited when generating pseudo-points (on the processed point-clouds generated by the LiDAR sensor) closer than the attacker position.
Spoofing attacks can also make an object appear closer or farther away, even beyond the range of the sensor, so that the actual object will not be considered as an obstacle.

In contrast to the replay attack of~\cite{Petit2015RemoteAO}, the \emph{white-box} and \emph{data-agnostic} attack of~\cite{Shin2017} can  \emph{add} pseudo-points (i.e.,~false positives) closer than the spoofer location.
An adversarial device receives the LIDAR pulse signal, adds the required delay, and fires the delayed laser pulse back to the LiDAR sensor. Moreover, a saturation attack can blind the sensors, thus \emph{hiding} objects by illuminating the LiDAR with a light of the same wavelength as that of the sensor.
A weak light source produces randomly-located pseudo-points, whereas a strong light source directed to the LiDAR completely blinds the sensor or a part of its field of view. In addition, a strong light source oblique to the LiDAR generates fake points in a direction other than that of the light of source.

Finally, the adversarial robustness to spoofing attacks on LiDAR sensors with deep neural network (DNN) detectors is studied in~\cite{Cao2019b}. A \emph{white-box} and \emph{data-agnostic} adversarial attack can strategically control the fake points to fool the machine learning model in the object detection step. The method \emph{adds} adversarial fake points at a specific location (distance, altitude, azimuth) with respect to the LiDAR sensor, with an attack success rate of 75\% in generating front-near false positives (5~meters in front of an AV). Note that, in this case, all spoofed objects are detected as vehicles, even though this was not in the specific objectives of the attack's formulation.

\subsection{Camera}
We can identify two types of \emph{gray-box} and \emph{data-agnostic} attacks that emit light~\cite{Petit2015RemoteAO}, namely,  \emph{camera blinding} and \emph{auto control confusion}. The blinding attack  shines  light on the camera sensor, thus \emph{hiding} any object within its field of view. Blinding, which can be full or partial, occurs when the camera is unable to tune down the auto-exposure or the gain, thus generating over-exposed images. The effectiveness of the blinding attack depends on the environmental light, the light source used to blind (i.e.,~wavelength), and the distance between the adversarial light source and the camera. Experiments in bright and dark environments with different light sources at multiple distances show that 650~nm (red) lasers can efficiently hide objects. Regarding auto control confusion, in which the goal of the attack is to influence the auto-exposure control of the camera, 
the attacker emits bursts of light toward the camera, and, during the stabilization to the new environmental conditions (typically between 1 and 6~seconds), the AV cannot detect objects. 
The attacker continuously switches the light on and off to confuse the auto controls, and, therefore, this attack differs from situations in which the camera can adapt more gradually to the new conditions, such as when driving out of a tunnel.
While the blinding attack could successfully blind -- in a simulated evaluation with low-quality images -- a 
widely used camera system at specific locations for up to 6~seconds when the attack is performed with burst lights from a 50~cm distance, this attack is only effective within a limited range (50-200~cm).
Under similar settings as~\cite{Petit2015RemoteAO}, the effect of different light sources on \emph{hiding} objects and other useful visual information is studied in~\cite{yan2016}.
Although attacking with infrared LED light does not affect the camera, laser beams pointed directly to the camera cause complete blindness for approximately 3~seconds.

\bigskip
\noindent To summarize, adversarial attacks with devices can cause significant problems in various tasks like automatic parking (ultrasound, radar), obstacle detection (LiDAR), or capturing of the surroundings (camera). However, they usually require the hardware specifications of the individual sensors to be known, while their effectiveness depends on the speed of the vehicles, the range of the attack, and the precision of the used devices.

\section{Attacks with objects and patches}
\label{subsec:object_based}

In this section, we discuss physical adversarial methods that use objects and patches to attack LiDARs and cameras.

\subsection{LiDAR attacks with physical objects}

Adversarial methods can generate physical 3D adversarial objects to mislead DNN-based LiDAR detection systems~\cite{Cao2019a}.
In the \emph{black-box} settings, a \emph{data-specific} evolutionary-based attack is used to hide a generated adversarial object that is placed in the environment. An iterative procedure is used to  generate the object: from an initial mesh of vertices,  a new population of vertices is generated at each iteration step by adding random perturbations drawn from a Gaussian distribution. However, when the object is placed in the environment, low attack success rates are obtained. 
A \emph{data-agnostic} method to \emph{hide} an object or to \emph{misclassify} a detected one is also proposed for \emph{white-box} settings. The objective of the method is to generate a synthetic adversarial object by perturbing the vertices of an original one, such that the AV system makes incorrect predictions.
For synthesizing the adversarial objects, a differentiable LiDAR renderer is first simulated; the feature aggregation is, then, formulated with a differentiable proxy function; and, finally, the smoothness of the generated adversarial object is ensured by devising different loss functions in the detection model.
In \emph{white-box} settings, the \emph{hiding} attack creates an adversarial object that is not detected 71\% of the time, while the \emph{misclassification} attack creates adversarial objects that are detected as another class (e.g.,~pedestrian) with high confidence. 

\subsection{Camera attacks with physical objects}
%
The first \emph{white-box} method for generating adversarial examples to \emph{misclassify} the decisions of an \emph{image classifier} even when transferred to the physical world was based on the expectation over transformation (EoT) framework~\cite{Athalye2017}. EoT  is used to construct APs that are robust to a set of physical transformations, such as random rotation, translation, additive noise, or 3D rendering of a texture. 
The generated 3D objects remain adversarial even when changing the viewpoint, translation and rotation of the object, lighting conditions, camera noise, and other physical world factors, with a classification accuracy dropping from 68.8\% to only 1.1\%.

Adversarial traffic signs that cause \emph{misclassification} can be generated with an  out-of-distribution attack or  a lenticular attack~\cite{Sitawarin18}.
An out-of-distribution adversarial example can be generated in \emph{white-box} and \emph{data-specific} settings, starting from an arbitrary non-traffic-sign image (i.e.,~a fast food chain logo), such that the arbitrary sign is classified as a traffic sign with high confidence. The AP is forced to lie only on the sign and not on the background, and to be invariant to different transformations, such as rotation, shearing, resizing, and randomized brightness. The lenticular attack relies on an optical phenomenon to deceive traffic sign DNN-based classifiers.
The idea is based on the fact that the camera will capture the images from a different height and angle than the passengers within the vehicle; thus, the AP might be perceived only from the camera's angle but not from the passengers' points of view. 

A \emph{white-box},  \emph{class-universal} (universal for any stop sign), \emph{object}-based attack may target  DNN-based classifiers by generating  visual APs that are robust to different physical conditions, such as different distance, viewpoint, illumination, and rotation~\cite{Eykholt2018}.
To ensure that such conditions exist in the image samples, real images were collected from the road, while synthetic ones were simulated using a set of transformations. To guarantee that the perturbations are only applied to the surface of the target object, a mask is used to project the computed perturbations to a physical region on the surface of the object (the road sign). In addition, the mask generates APs that are visible but inconspicuous to human observers (i.e.,~the mask looks like a graffiti). Also, since the position of the mask has an impact on the effectiveness of the attack, sparsity is enforced on the generated perturbations so that they are concentrated on regions where the system is more vulnerable. Finally, to account for fabrication errors, a term is added in the objective function to model color reproduction inaccuracies of the printing process.
The generated adversarial stop signs were evaluated on a vehicle, with an initial distance from the sign of 76~m, the speed of the vehicle varying up to 32~km/h, and the vehicle moving toward the sign. For two classifiers trained on different traffic sign datasets (achieving 91\% and 95.7\% classification accuracy), their accuracies dropped to zero.

A \emph{white-box} and \emph{class-universal} attack can generate  robust physical APs against a DNN for  \emph{object detection}~\cite{Chen18}, in the sense that the generated APs remain adversarial under various physical condition changes (e.g.,~distance, angle). This attack  focuses on the region proposal stage of the detector, which produces various candidate detection regions that are fed into the classifier stage. The method simultaneously attacks the classification within each proposed candidate region by approximately maximizing, through backpropagation, the classification adversarial loss for every proposed region. After  a forward pass on the region proposal network, the pruned region proposals are treated as constants for the classification. Then, the EoT framework is adopted for creating APs that are robust to physical transformations. Finally, the generated robust physical APs can be printed and applied on physical stop signs.
With the goal of \emph{misclassification} at the corresponding part of the model, the generated signs are printed and placed in an indoor scene with such distance and angle for mimicking driving conditions.
When the attack aims at a specific misclassification label, such as a person (targeted attack), the success rate is 87\%. When the attack does not target a specific label (untargeted attack), the success rate is 93\%. The robustness of the untargeted attack is mostly affected by the distance increment, whereas the robustness of the targeted attack is affected by the view angle. Although the attack focuses on the classification task in a drive-by evaluation (a vehicle approaches the sign from 60.9~meters away and with a speed between 8 and 24~km/h), the adversarial examples are also able to affect the detection. A clean stop sign is always detected and classified correctly, but the adversarial signs, on most occasions, are not detected; and when they are detected, they are misclassified most of the time.

Finally, based on~\cite{Eykholt2018}, an adversarial attack to object detectors is proposed in~\cite{Eykholt2018b} with the goal of \emph{hiding} (preventing the detection of) stop signs. By defining a loss function that outputs the maximum probability of a stop sign occurring within the scene, the extended \emph{white-box}, and \emph{class-universal} method minimizes this probability until it falls below the detection threshold of the classifier.
The method is evaluated indoors and outdoors, with recordings beginning at 9.1~meters away from a stop sign and ending when the sign is outside of the camera’s field of view. The attack success rate is 85.6\% indoors and 72.5\% outdoors, and, when the adversarial examples are evaluated on a different model, the attack success rate (transferability) for the indoor case remains high at 85.9\%, while, for the outdoor case, it degrades and drops to 40.2\%.

\subsection{Camera attacks with patches}
\label{subsec:patch}




Physical adversarial methods may also use a specific subcategory of objects, namely, patches, to attack cameras for an \emph{image classification} or an \emph{object detection} task. A \emph{data-agnostic}, targeted adversarial attack can generate  APs that  force a DNN to \emph{misclassify} objects by  localizing a single perturbation on every image such that the resulting patch can be printed and installed on objects~\cite{Brown2017}. The patch is the result of training over a variety of images, where, to encourage the trained patch to work regardless of the background, multiple transformations based on the EoT framework are applied on the patch in each image. Furthermore, an ensemble patch method is used in which a single patch is jointly crafted across five  models.
In simulated physical attacks (an adversarial patch is added to the image but not printed), for patches occupying only 10\% of the image size, the single patch achieves an attack success rate of around 89\%, while the ensemble patch leads to an attack success rate of around 95\% (averaging across all ensemble classifiers).

The patch-based method introduced in~\cite{Eykholt2018} (Section~\ref{subsec:object_based}), follows the same approach as the object-based one, with the difference that the AP is masked to take the form of black-and-white patches that are placed on traffic signs. The generated adversarial patches are evaluated on a vehicle with two classifiers, with the vehicle moving toward the sign from an  initial distance from the sign of 76~meters at a speed varying to up to 32~km/h. As the vehicle approaches the sign, the accuracies of both classifiers show a significant drop.

A \emph{white-box} and \emph{data-agnostic} physical patch generation method can  \emph{add} false-positive stop signs to the detector output~\cite{Eykholt2018b}. To this end, a composite loss function is used such that it first creates a new object localization, followed by a targeted misclassification.
A state-of-the-art detector identifies a nonexistent stop sign between 25\% and 79\% of the time when these patches are placed on a clean wall, where no actual stop sign exists, at 3~meters from the camera. 

Unlike attacks that focus on targets with no-intra class variety (i.e.,~only stop signs), 
target types with large intra-class variety (persons) are considered in~\cite{Thys19}. The loss function to help \emph{hide} people from detectors  considers three factors, namely, a non-printability score (how well the colors of a patch can be reproduced by a printer); the total variation of the image (favoring a patch with smooth color transitions); and the maximum objectiveness score in the image (i.e.,~the effectiveness on hiding a person), which aims to minimize the object or class score output by the detector. Operating in \emph{white-box} settings and constraining the patch to lie in the neighborhood of the predicted bounding boxes of persons, the method is able to generate a \emph{class-universal} adversarial patch for hiding persons.
This adversarial example patch decreases the recall of the detector to 26.4\% on a dataset of images in which people were always detected in clean conditions.

Finally, a patch-based attack for object detectors can potentially \emph{hide} all of the objects to be detected in a scene without overlapping with any of them~\cite{Lee2019}. The patch can be placed anywhere in the field of view of the camera and cause  all objects to disappear, even if they are far from the patch itself. This \emph{white-box} and \emph{data-agnostic} method generates the patch by maximizing the loss for the original targets, given a set of physical transformations that are applied on the patch using the EoT framework.
The patch can hide most of the objects within the field of view: the mean average precision drops from 40.9\% to 7.2\% when the patch is  randomly placed. The patch also disables the detection of objects that are moving, as long as the patch itself is static with respect to the camera. The patch is, to some extent, invariant to location, but it has weaker influence on objects that are farther away.

\bigskip
\noindent Overall, adversarial attacks with objects pose a very serious threat for AVs, in particular, for the common cases of attacking LiDAR and camera sensors that use DNN-based classifiers and detectors. These attacks can generate 3D adversarial objects that are placed in the physical world or even take the form of small 2D stickers (patches) that can be placed on any object. They have the ability to cause wrong classification or false detection, while they can also be used to hide all of the other objects that are around them (even pedestrians). However, despite their success, such attacks are usually applied in the white-box settings, and, generally, show low transferability to unknown (black-box) models.

\section{Countermeasures}
\label{sec:countermeasures}

In this section, we discuss countermeasures against adversarial physical attacks. We also discuss general methods, such as adversarial training (AT), that are applied to mitigate the effects of on-signal attacks, and their adaptation to physical attacks is straightforward.



\subsection{Ultrasonic and RADAR sensors}
A method against spoofing and jamming attacks can use each ultrasonic sensor to separate spoofed echoes from real ones and report the real distance (resilient obstacle detection)~\cite{Xu2018}. Ultrasonic sensors transmit pings of the same signal throughout their lifetimes and search for only the first echo. Since there is no bond between a ping and its echoes, to detect attacks and possibly reject spoofed echoes, a physical signal authentication is achieved by shifting the signal parameters through a challenge-response scheme: first, by customizing the ping signal, and, then, by correlating the received echoes with the pings. In addition, a set of ultrasonic sensors can also collaboratively detect an attack using a single-transmitter, multiple-receiver sensor structure for resilient obstacle detection and attacker localization~\cite{Xu2018}. These approaches can be also adapted for RADAR sensors.

\subsection{GPS}
Detection mechanisms may defend against attacks on the location, time, and Doppler shift parameters~\cite{Papadimitratos2008a}.
For a given parameter, the receiver first collects data during periods of time that are in normal mode (training data). Then, based on the normal-mode data, the receiver predicts the future values of the parameter and compares the predicted values with the ones obtained from the GNSS. If the difference with the predicted values exceeds a selected threshold, the receiver deems itself to be under attack, and all PVT solutions are discarded.

Two more spoofing countermeasures are described in~\cite{Humphreys2008}.
The first one focuses on the latency between the spoofing and original data bitstreams, where the receiver looks for a data bit sign change between consecutive accumulations at the coarse acquisition code-length interval. If a sign change is detected anywhere other than at an expected data bit boundary, the target receiver raises a flag. The second one focuses on detecting the vestigial authentic signal. The receiver copies the incoming digitized front-end data into a buffer used only for vestigial detection and selects one of the GPS signals being tracked, removes it from the data in the buffer, and performs acquisition for the same signal (the same pseudorandom noise identifier) on the buffered data. These steps are repeated for the same GPS signal, and the results are summed non-coherently until a probability of detection threshold is met. If a significant vestigial signal is present in the data, this technique will reveal it.

Finally, real-time GNSS spoofing detection can be achieved by processing beat carrier-phase measurements from an antenna pair~\cite{psiaki2014}. Considering a spoofed and a non-spoofed signal model, the observed differential beat carrier phases are fit to either of the two models, and the estimates (along with their associated fit error costs) are used to develop a spoofing hypothesis test. Then, real-time spoofing detection can be achieved through a switched-antenna version of the two-antenna system that determines a single-differenced beat carrier phase. Such a system can use  a radiofrequency (RF) switch between the two antennas, a single RF front-end, and a single receiver channel per tracked signal.

\subsection{LiDAR}
Countermeasures can be applied at the hardware or at the software level. At the hardware level, using multiple LiDAR sensors of different wavelengths, which collaboratively provide the scanned information, makes it harder for the attacker to manipulate every signal at the same time~\cite{Petit2015RemoteAO}. At the software level, an option is to (non)-predictably skip certain emitting pulses, which is similar to varying the scan speed. If the system notices a response that corresponds to a skipped pulse, then this is an indication of a possible attack. Another option is to shorten the pulse period, which consequently reduces the attack window. However,  lowering the pulse period  also shortens the  range of the sensor.

Defenses for DNNs can operate at the AV system, sensor, or machine learning model levels~\cite{Cao2019b}.
At the AV level, by studying the spoofed 3D point clouds, it can be observed that points from ground reflection are clustered into obstacles due to the information loss introduced in the pre-processing phase. Mapping a 3D point cloud into a 2D matrix results in height information loss, which facilitates an attack. To mitigate this effect, the ground reflection can be  filtered out in the pre-processing phase,  or, to reduce the information loss, the point cloud should not be transformed into an input feature matrix. At the sensor level, similarly to~\cite{Petit2015RemoteAO}, a possible defense consists of adding some randomness in the LiDAR pulses: by firing a random grouping of laser pulses at each cycle, an attacker would not know which reflections the sensor would be expecting. Finally, at the machine learning model level, an option is to perform adversarial training that, as we will discuss later, improves the robustness of DNN-based systems against adversarial attacks.

\subsection{Camera}
\label{subsec:vision_countermeasures}
Increasing the robustness of DNN-based vision systems against APs is a very challenging and widely studied problem. We focus here  on the tasks of \emph{image classification} and \emph{object detection}. 

Most defenses are for specific or weak attacks (and, thus, stronger attacks can bypass them~\cite{Carlini2017BreakD}) or just obscure the model (gradient masking), rather than making the classifier robust against all attacks~\cite{Athalye2018}. A defense that is empirically robust against all designed attacks is \emph{adversarial training} (AT). AT aims to \emph{train robust models} rather than defend against specific attacks: instead the model being trained only with the original (clean) data, it is trained along with the adversarial examples. AT has not been studied for physical APs yet, and  we, therefore,  provide here the results for on-signal attacks. Adaptation to physical APs is straightforward. 

Formally, for a signal $\x\in\Phi\sim\mathcal{X}$, we define a set of allowed perturbations $\mathcal{S}_{\v}\subseteq\Phi$ that formalizes the manipulative power of the attacker; that is, as described in Section~\ref{subsec:definitions}, the distortion caused by $g_{\v}$ shall not considerably change or destroy the information of the signal. Assuming data targets $y\in\mathcal{Y}$, a loss function $\mathcal{L}$, and model parameters $\boldsymbol{\theta}\in\mathbb{R}^p$, the goal of AT is to solve the following minimax problem~\cite{madry2017}:
\begin{equation}
\underset{\boldsymbol{\theta}}{\text{min}} \enskip \mathbb{E}_{(\x,y)\sim(\mathcal{X}, \mathcal{Y})}\big[\underset{\v\in\mathcal{S_{\v}}}{\text{max}}\enskip\mathcal{L}(g_{\v}(\x),y,\boldsymbol{\theta})\big],
\label{eq:adv_train}
\end{equation}
which corresponds to learning a model using signals (adversarial examples) that maximize the adversarial loss of the model.

For the simple case of the MNIST
image dataset~\cite{mnist}, AT is effective since the adversarial accuracy of the model (the accuracy of the model when given adversarial examples as inputs) remains close to 90\%, without a significant impact on the accuracy with clean examples, which is almost 98\%. For the slightly more complex CIFAR-10~\cite{cifar} image dataset though, the adversarial accuracy is almost 52\%, while the clean accuracy drops to almost 80\%. Robustness is, at the moment, at odds with accuracy~\cite{tsipras2019, fawzi2017_Analysis}: to increase robustness one might have to sacrifice accuracy. In addition, the cost of computing adversarial examples for the whole training procedure is very high, especially for high-dimensional datasets such as ImageNet~\cite{imagenet}.

However, as AT decreases the curvature of the decision boundaries of the classifier~\cite{Moosavi2019}, a regularization term can be added to the training loss to   impose smoothness. The minimization of this loss during training produces results comparable to AT on MNIST and CIFAR-10 but much faster. Moreover, a fast AT process~\cite{Shafahi2019} updates both the model parameters and image perturbations using one simultaneous backward pass, thus, with almost no additional cost to natural training, resulting in ImageNet models with around 44\% adversarial and 65\% clean accuracy.

For the task of \emph{object detection}, the authors in~\cite{Zhang2019} focus on the design principles of existing adversarial attacks; an attack to a detector can be achieved with variants of individual task losses or their combinations (localization and/or classification), and, in fact, adversarial attacks dedicated to one of the tasks can reduce the performance of the model on the other task.
Exploiting these interactions between different task losses, they generalize AT from classification to detection to form a new minimax training scheme, which combines task-oriented adversarial examples that maximize the adversarial loss of the detector. This  scheme improves by 20-30\% both the adversarial classification accuracy and the adversarial object detection accuracy, which is the ratio between the correctly detected objects and the actual total number of objects.

\section{Discussion and Outlook}

In this section, we summarize the findings from the evaluation of the adversarial attacks and countermeasures, and identify four main research directions toward designing robust sensing for 
autonomous driving systems.

\subsection{Evaluation}
The evaluation of adversarial attacks has been limited to specific sensors or to machine learning models. When the decision process is based on machine learning models, many attacks are evaluated on a few models or on similar architectures, and/or with models trained on the same or similar datasets. The  corresponding evaluation is, therefore, limited, as an attack that works well on a given dataset and architecture might not generalize. 

In fact, increasing the variability of the datasets and architectures is important to evaluate the transferability of the attack, since the generated adversarial examples might not transfer to models of different configurations. Also, most methods operate under white- or gray-box settings, which are unrealistic, as an attacker might not have (full) knowledge of the underlying sensory systems or the machine learning models. To satisfactorily evaluate the robustness of AVs to adversarial perturbations, it is desirable to design attacks in black-box settings that are invariant to physical transformations, such as rotations, changes in viewing angle, and illumination variations. Simulators provide a promising direction to test a wide variety of situations~\cite{Boloor2019}.


\subsection{Fusion}
While adversarial vulnerabilities of individual sensory systems have been thoroughly investigated, the effect on fused signals is still not studied. In adversarial settings, determining the modalities that contribute more to incorrect inferences could provide new insights on how the information provided by the non-adversarial sensors can support the systems and help them to remain robust ({\em adversarial robustness through signal fusion}). Moreover, from the perspective of the robustness of the fused signal, the consideration of \emph{multi-modal} adversarial methods (attacking the fusion itself) is an exciting research area that could open new directions in understanding vulnerabilities in adversarial settings. It is, therefore, desirable to study the effect of each individual sensor (under adversarial attacks) to the fusion-based decision-making process. 

\subsection{Adversarial Training}

AT has been mostly studied for image classification and object detection, especially for on-signal APs. AT is a general framework that can also be exploited to achieve robustness against \emph{any} type of AP (e.g.,~patchbased), including physical APs, by replacing on-signal adversarial examples with physical adversarial examples. Moreover, the  principles of AT are applicable to \emph{any} sensing modality when the underlying system is a machine learning model. However, current versions of AT schemes might saturate to some sub-optimal solution, which could explain the increasing gap between adversarial and standard accuracy~\cite{tsipras2019}. Finally, it is also important to focus on the dynamics of AT and explore \emph{how} and \emph{why} it improves adversarial robustness~\cite{Moosavi2019}. 

\subsection{Robustness}
When evaluating the robustness of deep models against adversarial attacks, it is not only important to ensure that the selected attacks are strong (to avoid overfitting to weak attacks~\cite{Carlini2017BreakD}), but also to test the models using gradient-free black-box attacks~\cite{uesato18a} to ensure that the models are actually robust and that the gradients of the models are not just obfuscated (gradient masking), which gives a false sense of robustness~\cite{Athalye2018}.

Ultimately, a key question is to determine why APs exist. Their existence is expected to be due to a combination of causes related to the data, the architecture, and the training/learning scheme. With AT, robustness is an effect of the modified data and learning scheme. When studying the architectures, increasing the capacity of the model seems to help AT to increase robustness~\cite{madry2017}. Important open questions include the following. What would be the effect of other network elements (e.g.,~the size of the learned filters)? What are the interactions of all of these entities that cause (or can prevent) such adversarial vulnerabilities? In that sense, for building and deploying safer systems, instead of black-box ones, more transparent models that allow us to interpret and explain their functionalities and reasoning are certainly desirable. Understanding how these systems work might, indeed, shed light on the underlying causes of their adversarial vulnerabilities.

\section*{Acknowledgments}

This work is supported by the CHIST-ERA program through the project CORSMAL, under UK EPSRC grant EP/S031715/1 and Swiss NSF grant 20CH21{\_}180444.

\bibliographystyle{IEEEtran}
\bibliography{refs}

\begin{thebibliography}{10}
\providecommand{\url}[1]{#1}
\csname url@samestyle\endcsname
\providecommand{\newblock}{\relax}
\providecommand{\bibinfo}[2]{#2}
\providecommand{\BIBentrySTDinterwordspacing}{\spaceskip=0pt\relax}
\providecommand{\BIBentryALTinterwordstretchfactor}{4}
\providecommand{\BIBentryALTinterwordspacing}{\spaceskip=\fontdimen2\font plus
\BIBentryALTinterwordstretchfactor\fontdimen3\font minus
  \fontdimen4\font\relax}
\providecommand{\BIBforeignlanguage}[2]{{%
\expandafter\ifx\csname l@#1\endcsname\relax
\typeout{** WARNING: IEEEtran.bst: No hyphenation pattern has been}%
\typeout{** loaded for the language `#1'. Using the pattern for}%
\typeout{** the default language instead.}%
\else
\language=\csname l@#1\endcsname
\fi
#2}}
\providecommand{\BIBdecl}{\relax}
\BIBdecl

\bibitem{Khan2018}
A.~{Khan}, B.~{Rinner}, and A.~{Cavallaro}, ``{Cooperative Robots to Observe
  Moving Targets: A Review},'' \emph{IEEE Transactions on Cybernetics},
  vol.~48, no.~1, pp. 187--198, 2018.

\bibitem{Eykholt2018}
K.~Eykholt, I.~Evtimov, E.~Fernandes, B.~Li, A.~Rahmati, C.~Xiao, A.~Prakash,
  T.~Kohno, and D.~Song, ``{Robust Physical-World Attacks on Deep Learning
  Visual Classification},'' in \emph{Proc. IEEE on Computer Vision and Pattern
  Recognition}, Salt Lake City, UT, Jun. 2018, pp. 1625--1634.

\bibitem{Fawzi2016}
A.~Fawzi, S.-M. Moosavi-Dezfooli, and P.~Frossard, ``{Robustness of
  classifiers: from adversarial to random noise},'' in \emph{Advances in Neural
  Information Processing Systems}, Barcelona, Spain, Dec. 2016, pp. 1632--1640.

\bibitem{Fawzi2017}
A.~{Fawzi}, S.-M. {Moosavi-Dezfooli}, and P.~{Frossard}, ``{The Robustness of
  Deep Networks: A Geometrical Perspective},'' \emph{IEEE Signal Processing
  Magazine}, vol.~34, no.~6, pp. 50--62, 2017.

\bibitem{ortiz-jimenezHoldMeTight2020}
G.~{Ortiz-Jimenez}, A.~{Modas}, S.-M. {Moosavi-Dezfooli}, and P.~Frossard,
  ``Hold me tight! {{Influence}} of discriminative features on deep network
  boundaries,'' \emph{arXiv:2002.06349}, Feb. 2020.

\bibitem{Yuan2019}
X.~{Yuan}, P.~{He}, Q.~{Zhu}, and X.~{Li}, ``Adversarial examples: Attacks and
  defenses for deep learning,'' \emph{IEEE Trans. on Neural Networks and
  Learning Systems}, vol.~30, no.~9, pp. 2805--2824, Sep. 2019.

\bibitem{yan2016}
C.~Yan, W.~Xu, and J.~Liu, ``{Can You Trust Autonomous Vehicles: Contactless
  Attacks against Sensors of Self-driving Vehicle},'' in \emph{DEF CON}, Las
  Vegas, NV, Aug. 2016, pp. 1--49.

\bibitem{Cao2019a}
Y.~Cao, C.~Xiao, D.~Yang, J.~Fang, R.~Yang, M.~Liu, and B.~Li, ``{Adversarial
  Objects Against LiDAR-Based Autonomous Driving Systems},'' 2019, preprint
  arXiv:1907.05418.

\bibitem{Eykholt2018b}
K.~Eykholt, I.~Evtimov, E.~Fernandes, B.~Li, A.~Rahmati, F.~Tram{\`e}r,
  A.~Prakash, T.~Kohno, and D.~Song, ``{Physical Adversarial Examples for
  Object Detectors},'' in \emph{USENIX Workshop on Offensive Technologies},
  Baltimore, MD, Aug. 2018.

\bibitem{Petit2015RemoteAO}
J.~Petit, B.~Stottelaar, and M.~Feiri, ``{Remote Attacks on Automated Vehicles
  Sensors: Experiments on Camera and LiDAR},'' in \emph{Black Hat Europe},
  Amsterdam, The Netherlands, Nov. 2015.

\bibitem{Sitawarin18}
C.~Sitawarin, A.~N. Bhagoji, A.~Mosenia, M.~Chiang, and P.~Mittal, ``{DARTS:
  Deceiving Autonomous Cars with Toxic Signs},'' 2018, preprint
  arXiv:1802.06430.

\bibitem{Seyed-Mohsen16}
S.-M. Moosavi-Dezfooli, A.~Fawzi, O.~Fawzi, and P.~Frossard, ``Universal
  adversarial perturbations,'' in \emph{Proc. IEEE on Computer Vision and
  Pattern Recognition}, Honolulu, HI, Jul. 2017.

\bibitem{Thys19}
S.~Thys, W.~{van~Ranst}, and T.~Goedeme, ``Fooling automated surveillance
  cameras: adversarial patches to attack person detection,'' 2019, preprint
  arXiv:1904.08653.

\bibitem{Papadimitratos2008b}
P.~{Papadimitratos} and A.~{Jovanovic}, ``{Protection and fundamental
  vulnerability of GNSS},'' in \emph{Proc. IEEE International Workshop on
  Satellite and Space Communications}, Toulouse, France, Oct. 2008.

\bibitem{Xu2018}
W.~{Xu}, C.~{Yan}, W.~{Jia}, X.~{Ji}, and J.~{Liu}, ``{Analyzing and Enhancing
  the Security of Ultrasonic Sensors for Autonomous Vehicles},'' \emph{IEEE
  Internet of Things Journal}, vol.~5, no.~6, pp. 5015--5029, 2018.

\bibitem{Humphreys2008}
T.~Humphreys, B.~Ledvina, M.~Psiaki, B.~O'Hanlon, and P.~Kintner, ``{Assessing
  the Spoofing Threat: Development of a Portable GPS Civilian Spoofer},'' in
  \emph{ION GNSS Conference}, Savannah, GA, Sep. 2008.

\bibitem{Shin2017}
H.~Shin, D.~Kim, Y.~Kwon, and Y.~Kim, ``{Illusion and Dazzle: Adversarial
  Optical Channel Exploits Against Lidars for Automotive Applications},''
  \emph{Cryptographic Hardware and Embedded Systems}, pp. 445--467, 2017.

\bibitem{Cao2019b}
Y.~{Cao}, C.~{Xiao}, B.~{Cyr}, Y.~{Zhou}, W.~{Park}, S.~{Rampazzi}, Q.~A.
  {Chen}, K.~{Fu}, and Z.~{Morley Mao}, ``{Adversarial Sensor Attack on
  LiDAR-based Perception in Autonomous Driving},'' 2019, preprint
  arXiv:1907.06826.

\bibitem{Athalye2017}
A.~{Athalye}, L.~{Engstrom}, A.~{Ilyas}, and K.~{Kwok}, ``{Synthesizing Robust
  Adversarial Examples},'' 2017, preprint arXiv:1707.07397.

\bibitem{Chen18}
S.-T. Chen, C.~Cornelius, J.~Martin, and D.~H. Chau, ``{ShapeShifter: Robust
  Physical Adversarial Attack on Faster R-CNN Object Detector},'' \emph{arXiv
  e-prints}, 2018, preprint arXiv:1804.05810.

\bibitem{Brown2017}
T.~B. Brown, D.~Man{\'e}, A.~Roy, M.~Abadi, and J.~Gilmer, ``Adversarial
  patch,'' in \emph{Advances in Neural Information Processing Systems}, Long
  Beach, CA, Dec. 2017.

\bibitem{Lee2019}
M.~{Lee} and Z.~{Kolter}, ``{On Physical Adversarial Patches for Object
  Detection},'' 2019, preprint arXiv:1906.11897.

\bibitem{Papadimitratos2008a}
P.~Papadimitratos and A.~Jovanovic, ``{GNSS-based} positioning: Attacks and
  countermeasures,'' in \emph{Proc. IEEE Conference on Military
  Communications}, San Diego, USA, Nov. 2008.

\bibitem{psiaki2014}
M.~Psiaki, B.~O'Hanlon, S.~Powell, J.~Bhatti, K.~Wesson, T.~Humphreys, and
  A.~Schofield, ``{GNSS spoofing detection using two-antenna differential
  carrier phase},'' \emph{ION GNSS Conference}, vol.~4, pp. 2776--2800, Jan.
  2014.

\bibitem{Carlini2017BreakD}
N.~Carlini and D.~Wagner, ``{Adversarial Examples Are Not Easily Detected:
  Bypassing Ten Detection Methods},'' in \emph{ACM Workshop on Artificial
  Intelligence and Security}, Texas, USA, Nov. 2017.

\bibitem{Athalye2018}
A.~Athalye, N.~Carlini, and D.~Wagner, ``Obfuscated gradients give a false
  sense of security: Circumventing defenses to adversarial examples,'' 2018,
  preprint arXiv:1802.00420.

\bibitem{madry2017}
A.~Madry, A.~Makelov, L.~Schmidt, D.~Tsipras, and A.~Vladu, ``{Towards Deep
  Learning Models Resistant to Adversarial Attacks},'' in \emph{Proc.
  International Conference on Learning Representation}, Vancouver, Canada, Apr.
  2018.

\bibitem{mnist}
Y.~LeCun and C.~Cortes, ``Mnist handwritten digits database,'' \emph{Dataset
  available from http://yann.lecun.com/exdb/mnist/}, 2010.

\bibitem{cifar}
A.~Krizhevsky, ``Learning multiple layers of features from tiny images,''
  University of Toronto, Tech. Rep., 2009.

\bibitem{tsipras2019}
D.~Tsipras, S.~Santurkar, L.~Engstrom, A.~Turner, and A.~Madry, ``{Robustness
  May Be at Odds with Accuracy},'' \emph{Proc. International Conference on
  Learning Representation}, May 2019.

\bibitem{fawzi2017_Analysis}
A.~Fawzi, O.~Fawzi, and P.~Frossard, ``Analysis of classifiers' robustness to
  adversarial perturbations,'' \emph{Machine Learning}, vol. 107, no.~3, pp.
  481--508, March 2018.

\bibitem{imagenet}
J.~Deng, W.~Dong, R.~Socher, L.~J. Li, L.~Kai, and F.~F. Li, ``{{ImageNet}}:
  {{A}} large-scale hierarchical image database,'' in \emph{{{IEEE Conference}}
  on {{Computer Vision}} and {{Pattern Recognition}} ({{CVPR 2009}})}.\hskip
  1em plus 0.5em minus 0.4em\relax Miami Beach, FL: {IEEE}, 2009, pp. 248--255.

\bibitem{Moosavi2019}
S.-M. Moosavi-Dezfooli, A.~Fawzi, J.~Uesato, and P.~Frossard, ``{Robustness via
  Curvature Regularization, and Vice Versa},'' in \emph{Proc. IEEE on Computer
  Vision and Pattern Recognition}, Long Beach, CA, Jun. 2019.

\bibitem{Shafahi2019}
A.~{Shafahi}, M.~{Najibi}, A.~{Ghiasi}, Z.~{Xu}, J.~{Dickerson}, C.~{Studer},
  L.~S. {Davis}, G.~{Taylor}, and T.~{Goldstein}, ``Adversarial training for
  free!'' in \emph{Advances in Neural Information Processing Systems},
  Vancouver, Canada, Dec. 2019.

\bibitem{Zhang2019}
H.~{Zhang} and J.~{Wang}, ``{Towards Adversarially Robust Object Detection},''
  in \emph{Proc. IEEE International Conference Computer Vision}, Seoul, South
  Korea, Oct. 2019.

\bibitem{Boloor2019}
A.~Boloor, X.~He, C.~D. Gill, Y.~Vorobeychik, and X.~Zhang, ``{Simple Physical
  Adversarial Examples against End-to-End Autonomous Driving Models},'' 2019,
  preprint arXiv:1903.05157.

\bibitem{uesato18a}
J.~Uesato, B.~O'Donoghue, P.~Kohli, and A.~{van den Oord}, ``Adversarial risk
  and the dangers of evaluating against weak attacks,'' in \emph{Proc. Int.
  Conf. on Machine Learning}, Stockholm, Sweden, Jul. 2018.

\end{thebibliography}

\end{document}